\documentstyle[tighten,prb,epsf,floats,aps]{revtex}
\date{June 10, 1997, submitted to Phys. Rev. B}
\begin{document}
%\draft
%\preprint{\today}
\twocolumn[
\hsize\textwidth\columnwidth\hsize\csname @twocolumnfalse\endcsname

\title{ Doping Dependence of the Pseudogap State in the $ab$-plane IR 
Response of La$_{2-x}$Sr$_{x}$CuO$_{4}$} 

\author{T.~Startseva$^{(1)}$, T.~Timusk$^{(1)}$, 
A.V.~Puchkov$^{(2)}$, D.N.~Basov$^{(3)}$, H.A.~Mook$^{(4)}$, 
T.~Kimura$^{(5)}$, and  K.~Kishio$^{(5)}$} \address{$^{(1)}$ 
Department of Physics and Astronomy,  McMaster University 
 \protect \\ Hamilton, Ontario, CANADA L8S 4M1}
\address{$^{(2)}$ Department of Applied Physics, Stanford University, 
Stanford, CA 94305} \address{$^{(3)}$ Department of Physics, 
University of California San-Diego, La Jolla, CA 92093} 
\address{$^{(4)}$ Oak Ridge National Laboratory, Oak Ridge, Tennessee 
37831} \address{$^{(5)}$ Department of Applied Chemistry, University 
of Tokyo, Tokyo 113, Japan} % 
 
\maketitle
%
% Abstract
%
\begin{abstract}

  The {\em ab}-plane optical spectra of two single crystals of La$_{2-
x}$Sr$_{x}$CuO$_{4}$, one underdoped  and one overdoped 
were investigated. We observe a gap-like depression  of the effective 
scattering rate $1/\tau(\omega, T)$ below 700 cm$^{-1}$ in both 
systems.  This 
feature  persists  up to 300~K in the underdoped sample with the 
concentration of Sr x=0.14 but loses  prominance at  temperatures above 
300~K in the overdoped regime (x=0.22). Below 700 cm$^{-1}$ 
$1/\tau(\omega, T)$ is temperature dependent and superlinear in 
frequency for both samples. Above this frequency the effective  
scattering rate becomes linear in frequency and is temperature 
independent in the case of the underdoped 
La$_{1.86}$Sr$_{0.14}$CuO$_4$ up to 300~K. On the other hand,  
 the overdoped  
La$_{1.78}$Sr$_{0.22}$CuO$_4$ shows a $1/\tau(\omega, T)$ 
temperature dependence above 700~cm$^{-1}$ at all temperatures. This 
behaviour of the frequency and temperature dependent scattering rates 
is a signature of a pseudogap state in other materials and suggests 
that both the under and overdoped single-layer HTSC systems 
La$_{2-x}$Sr$_{x}$CuO$_4$ have a pseudogap at temperatures exceeding 
300~K. 

\end{abstract} 
\narrowtext                                                 
]

Right from  the discovery of high-temperature superconductivity 
(HTSC) in the complex copper oxides, it has been recognized that the 
CuO$_2$ planes play an important role in the nature of this 
phenomenon. While all of the  cuprates share  this structural element, 
La$_{2-x}$Sr$_{x}$CuO$_4$ (LSCO)  possesses only one CuO$_2$ plane 
per unit cell which makes it an excellent prototype system for 
examining the role played by the CuO$_2$ planes. It is also a 
good model for the study of doping dependence since 
it can doped by the 
addition of strontium over a wide range: from the underdoped,  
where $T_c$ increases with Sr content, to the optimally doped where 
$T_c$ reaches its maximum value of $\approx 40$ K, and  to the overdoped 
region where $T_c\rightarrow 0 $ at $x=0.34$.\cite{batlogg} 

It is well known that the in-plane transport properties of HTSC 
materials are anomalous. For example, the in-plane dc resistivity, 
$\rho_{ab}(T)$,  for the samples doped close to the optimal doping 
level, is linear in temperature from the superconducting transition 
temperature $T_c$ to well above 900~K. At the same time, underdoped 
samples show a crossover from the linear $T$ dependence to a 
superlinear, $\rho_{ab}(T)=T^{1+\delta}$,  below a characteristic 
temperature $T^{\ast}$. It was shown by  B.~Batlogg {\it et al.} 
\cite{batlogg} that $T^{\ast}$ decreases from 800~K to approximately 
300~K as the doping level is increased from the strongly underdoped 
to just over the optimal doping level. Similar behavior at 
$T=T^{\ast}$ has been observed in the Hall effect coefficient and the 
magnetic susceptibility.\cite{Hwang,hall effect} 

\begin{figure}[t]
\leavevmode  
%\epsysize=0.6
%\epsfbox{6in}
\epsfxsize=0.75\columnwidth
\centerline{\epsffile{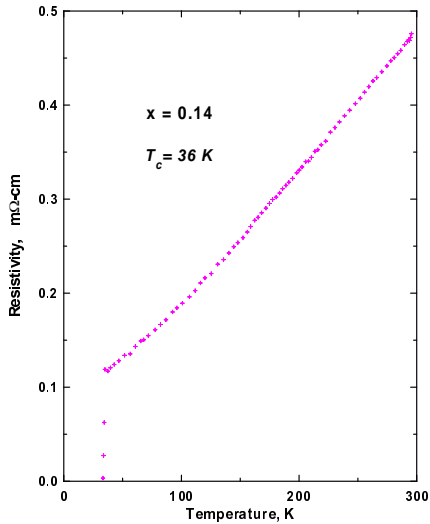}}
\vspace{0.1in}
\caption{The temperature dependence of the in-plane resistivity of
La$_{1.86}$Sr$_{0.14}$CuO$_4$ is shown with a sharp superconducting
transition at ~36~K. The shape of the curve is consistent with
$T^{\ast}$ being greater than 300~K.
}
\label{resistivity}
\end{figure}

A completely satisfactory microscopic understanding of the peculiar 
transport properties of HTSC is missing at this time. However, the 
temperature and doping dependences of the crossover behaviour at 
$T=T^{\ast}$ noted in the dc transport properties are in accord with 
the idea that a pseudogap forms in the spectrum of the low-energy  
electronic excitations responsible for the scattering of the charge 
carriers. A similar gap-like feature has been observed experimentally 
in the nuclear magnetic resonance (NMR)\cite{millis}, the angular 
resolved photoemission (ARPES)\cite{arpes}, specific 
heat\cite{specific heat} and the infrared 
optical\cite{basov96,puchkov96a,puchkov96b} (IR) measurements. All  of
these observations can be consistently interpreted in terms of a 
pseudogap. In this picture the dc transport properties and the gap 
formation are intimately related: a gap in the density of states near 
the Fermi level will result in a reduced 1/$\tau(\omega,T)$ in the 
temperature and spectral region where this gap occurs. Since the 
ARPES results suggest that the gap has a d$_{x^2-y^2}$ momentum 
dependence, momentum-averaging measurement techniques, like NMR or IR  
optical, would observe a {\it pseudogap} rather than a full gap even 
at lowest temperatures where this gap is fully formed. 

A  pseudogap feature was observed in the {\it c-axis} IR {\it 
conductivity} in YBa$_2$Cu$_3$O$_{7-x}$ (Y123) and YBa$_2$Cu$_4$O$_8$ 
(Y124) materials.\cite{basov96,homes} The $ab$-plane results show 
increased coherence (a narrower Drude peak) upon entering the 
pseudogap state while the  $c$-axis optical response shows a 
depressed conductivity.  The pseudogap state  is seen in both the 
$c$-axis conductivity and the $ab$-plane scattering rate in the same 
doping and temperature region, suggesting that the two phenomena are 
closely related.\cite{basov96} Recent $c$-axis optical results on 
single crystals of of  slightly underdoped 
La$_{1.86}$Sr$_{0.14}$CuO$_4$\cite{c-axis LSCO} show that  the 
pseudogap state in the $c$-axis direction of this material is not as 
well defined as it is in the two plane materials. However, as the 
doping is reduced further the $c$-axis pseudogap state features 
become clearer below 0.1 eV.\cite{uchida c-axis} 

The weak pseudogap as seen by  NMR and neutron scattering\cite{mason} 
in LSCO has led to the suggestion that  the existence of the 
pseudogap in the spin excitation spectrum is only possible in bilayer 
compounds such as Y123 and Y124. In particular, Millis and Monien 
attribute the pseudogap  (or the spin gap) to strong 
antiferromagnetic correlations between the planes in the bilayer, 
which are responsible for a quantum order-disorder transition.\cite{millis} 
However, the characteristic deviations below the linear 
extrapolation and $T^{\ast}$ seen in dc conductivity in the 
bilayer\cite{uchida} materials are also seen in LSCO.\cite{uchida} Thus, it is 
important to see if the characteristic depression of the frequency 
dependent scattering rate in the pseudogap state, seen in the 
bilayer materials\cite{basov96,puchkov96a,puchkov96b}, can also be 
observed in the single plane materials such as LSCO. 
 
\begin{figure}[t]
\leavevmode
\epsfxsize=0.75\columnwidth
\centerline{\epsffile{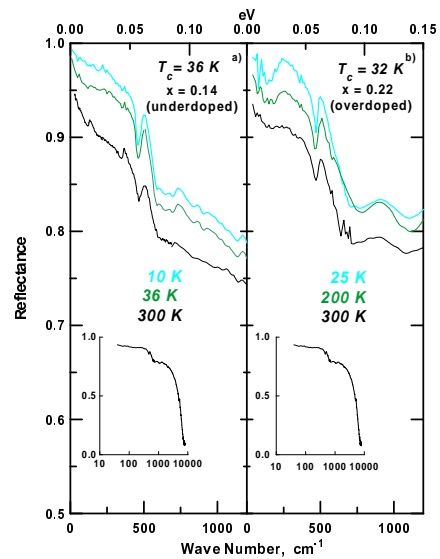}}
\vspace{0.1in}

\caption{The reflectance of
La$_{1.86}$Sr$_{0.14}$CuO$_4$  (a) and
 La$_{1.78}$Sr$_{0.22}$CuO$_4$ (b) is shown at three
temperatures:
10~K, 36~K and 300~K.
 The thickness of the line goes up as the temperature goes down.
Phonon peaks are visible at 150, 300 and 500~cm$^{-1}$. The insert in
the left panel is a semi-log graph of the reflectance at 300~K which
shows a  plasma edge around 7000~cm$^{-1}$.
}
\label{reflectivity}
\end{figure}

Previous work on the inplane far infrared  optical properties of the 
single layer lanthanum strontium cuprate includes work on the oxygen 
doped La$_2$CuO$_{4-\delta}$\cite{Quijada}, thin films of LSCO\cite{Gao} as 
well as single crystal work at room temperature\cite{uchida}. To our 
knowledge, a study of the temperature  and doping dependence has not 
been done. We fill this gap here by performing optical measurements 
on high-quality LSCO single crystals at temperatures ranging from 
10~K to 300~K at two different doping levels. 

To better display the effect of increased coherence resulting from 
the formation of the pseudogap state, a memory function, or 
extended Drude analysis is used.  In this treatment the complex optical 
conductivity is modeled by a Drude peak with a frequency-dependent 
scattering rate and an effective electron mass.\cite{goetze,allen} While 
the optical conductivity tends to emphasize free particle behaviour, 
a study of the frequency dependence of the effective scattering rate 
puts more weight on displaying the interactions of the free particles 
with the elementary excitations of the system.\cite{gold} 
The temperature evolution of the frequency dependent scattering rate 
and effective mass spectra are of particular interest and are  
defined as follows: 

\begin{equation} 
1/\tau(\omega, T) = {\omega^{2}_p\over4\pi} Re({1\over\sigma(\omega, T)})
\end{equation}
\begin{equation} 
{m^{\ast}(\omega, T)\over m_e} = {1\over 
\omega}{\omega^{2}_p\over4\pi} Im({1\over\sigma(\omega, T)}) 
\end{equation} 

Here, $\sigma(\omega, T)=\sigma_{1}(\omega ,T)+i\sigma_{2}(\omega, T)$ is 
the complex optical conductivity and $\omega_p$ is the plasma 
frequency of charge carriers.

The single crystals of La$_{2-x}$Sr$_x$CuO$_4$ with approximate 
dimensions 5x3x3~mm$^3$ were grown by the travelling-solvent floating 
zone technique at Oak Ridge \cite{mook} in the case of $x 
= 0.14$ and in Tokyo \cite{kimura} in the case of $x 
= 0.22$.   The critical temperature was determined by both SQUID 
magnetization and resistivity measurements and was found to be 
 36~K for the nominal concentration of Sr $x = 0.14$ and 32~K for 
$x = 0.22$. Since the highest $T_c$ in the LSCO system has been found 
at to be 40~K $x=0.17$, we conclude  that the $x = 0.14$ crystal is  
underdoped and the $x = 0.22$ is  overdoped.   The crystal with $x 
=0.14$ was aligned using Laue diffraction and polished parallel to 
the  CuO$_2$ planes. The crystal with $x = 0.22$ was polished in 
Tokyo in the direction of the ab-plane. It is  important to have the 
sample surface accurately parallel to the {\it ab}-plane to  avoid 
any c-axis contribution to the optical conductivity.\cite{miscut}  
The miscut of the sample off the ab-plane was checked by a high 
precision triple axis x-ray diffractometer  and was determined to be 
less than 0.8\%. 

All reflectivity measurements were performed with a Michelson 
interferometer using three different detectors which cover frequencies  
ranging from 10 to  10000 cm$^{-1}$. The experimental uncertainty in 
the reflectance data does not exceed 1$\%$. The dc resistivity 
measurements were carried out using a standard 4-probe technique.

The result of the resistivity measurement on the 
La$_{1.86}$Sr$_{0.14}$CuO$_4$  single crystal, used in the optical 
measurements, is shown in Fig.~1.
It is commonly accepted that the dc resistivity is linear at high 
temperatures for LSCO and that the pseudogap begins to 
form near the temperature where the resistivity drops below this 
linear trend.\cite{batlogg} At lower temperatures there is a region 
of superliner temperature dependent resistivity. The $T^{\ast}$ value 
for our samples with $x=0.14$ and $x=0.22$, extracted from the phase 
diagram of Batlogg {\it et al.},\cite{batlogg} are 450~K and 200~K, 
respectively. In agreement with this, the   
resistivity shows a superlinear  temperature dependence below  
room temperature as expected in the pseudogap region.

In Fig.~2  we present the reflectivity data at temperatures above and 
below $T_c$. For clarity, only three temperatures are shown: 
$T=300$~K, a temperature just above the superconducting  transition 
and a low temperature $\approx 10$~K or 25~K for $x=0.14$ and 
$x=0.22$, respectively, in the superconducting state.  In the 
frequency region shown the reflectance  is strongly temperature 
dependent for both materials, dropping  by approximately  $10\%$ as 
temperature is increased from the lowest temperature to $T=300$~K. 
The plasma edge is observed at 7800~cm$^{-1}$ (see insert of Fig.~2). 
The distinct  peaks at approximately 150, 300 and 500~cm$^{-1}$  in 
the  LSCO  reflectivity spectra correspond to the excitation of 
ab-~ plane $TO$ phonons.\cite{tajima_phonons}

The complex optical  conductivity $\sigma(\omega)$  was obtained by 
Kramers-Kronig analysis of the reflectivity data. Since, in principle, this 
analysis requires knowledge of the reflectance at all 
frequencies,  reflectivity extensions must be  used at high and low 
frequencies. The Hagen-Rubens formula was used for the low frequency 
reflectivity extrapolation, with parameters taken from the dc 
resistivity measurements on the same sample shown in Fig.~1 and the 
results  of H.~Takagi {\em et al.}\cite{takagi} for the overdoped 
sample. For the high-frequency extension with $\omega>8000$~cm$^{-1}$ 
we used reflectivity results of Uchida {\it et al.}\cite{uchida} At  
frequencies higher than 40~eV the reflectivity was assumed to fall as 
$1/\omega ^{4}$. 

We calculate the plasma frequency of the superconducting charge 
carriers and the London penetration depth using the following 
formula:\cite{Timusk review} 
\begin{equation} \epsilon_1 = 1 - 
{\omega_{ps}^{2}\over\omega^2}. 
\end{equation} 

\begin{figure}[t]
\leavevmode
\epsfxsize=0.95\columnwidth
\centerline{\epsffile{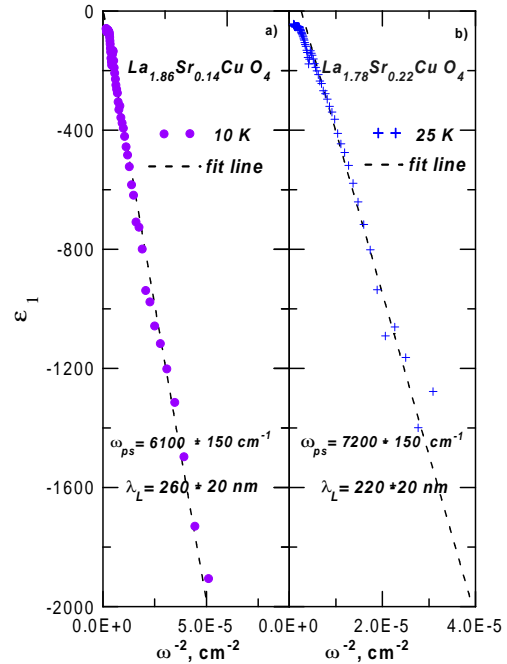}}
\vspace{0.1in}
 
\caption{The real part of the dielectric function as a function of
$\omega^{-2}$ of La$_{1.86}$Sr$_{0.14}$CuO$_4$ at 10~K in the panel
a) and of La$_{1.78}$Sr$_{0.22}$CuO$_4$ at 25~K at the panel b). The
dash line is linear fit. The slope of the fit gives the values of
both a superconducting plasma frequency of LSCO
 and  the London penetration depth.}
\label{epsilon}
\end{figure}

The slope of the low-frequency  dielectric function, 
$\epsilon_1(\omega)$, plotted  as a function of $w^{-2}$ in Fig.~3a,b  
gives  plasma frequencies of 6100~cm$^{-1}$ and 7200~cm$^{-1}$ 
in the superconducting state. The corresponding London penetration 
depths are $\lambda_L=1/2\pi\omega_{ps}=250$~nm and 220~nm for 
La$_{1.86}$Sr$_{0.14}$CuO$_4$ and La$_{1.78}$Sr$_{0.22}$CuO$_4$, 
respectively. These values  are in  good agreement with those 
obtained previously by Gao {\it et al.}  in films\cite{Gao}   
($\lambda_L = 275\pm5~nm$)  and by muon-spin-relaxation\cite{msr}  
($\lambda_L = 250~nm$).

The real part of the conductivity for the two materials is  shown in 
Fig.~4a,b. For completeness we also show the imaginary part of the 
conductivity in Fig.~4c,d. The real part of the conductivity has a Drude 
peak which narrows as the temperature decreases in agreement with the 
metallic temperature dependence of the dc resistivity. The 
conductivity of LSCO is temperature-dependent in the MIR frequency 
region as well.  This  temperature dependence becomes stronger as 
$T_{c}$ is reduced by overdoping. There are strong deviations 
from the Drude shape in the form of an onset or a step in the conductivity
at 700~cm$^{-1}$.  Unlike the Y123, Y124, and 
Bi$_2$Sr$_2$CaCu$_2$O$_{8+\delta}$ (Bi2212) materials which show 
similar features only at low temperature, the optical 
conductivity of La$_{1.86}$Sr$_{0.14}$CuO$_4$ shows a threshold at 
about 700~cm$^{-1}$ already at room temperature. 

Another deviation from the Drude form is a shift of the Drude peak 
from zero frequency to 150~cm$^{-1}$. This peak  grows in 
magnitude and narrows as  the doping level of Sr increases. A similar 
peculiarity was observed in the conductivity of a single crystal of 
La$_2$CuO$_{4+\delta}$ \cite{Quijada}; however, it was absent in the optical 
data of La$_{2-x}$Sr$_{x}$CuO$_4$ thin films.\cite{Gao} The nature of the peak 
is unclear. It is seen in many HTSC systems and has been attributed 
to localization.\cite{timusk95} An artifact of the polished surface 
is another possible explanation.

We define an overall plasma frequency in terms of the sum rule 
$w_p^2/8 = \int^{\omega_{max}}_{0} \sigma_1(\omega)d\omega $  with 
$\omega_{max}$ = 8000 cm$^{-1}$.  At room temperature, 
$w_p$ = 15100~cm$^{-1}$ and 13800~cm$^{-1}$ for LSCO with $x=0.14$ and 
$x=0.22$, respectively. These numbers are similar to those measured 
previously\cite{Gao,Tamasaku94,Quijada} and were used to calculate 
the frequency-dependent scattering rate using Eq.~1.  In a  recent 
survey on a number of compounds,  Puchkov {\it et 
al.}\cite{puchkov96b}  found that while the plasma frequency grows 
with doping in the underdoped region, this growth stops at optimal 
doping.  In agreement, we observe here a slight drop in the spectral 
weight as one moves from the slightly underdoped to the overdoped 
region.  

\begin{figure*}[t]
\leavevmode
\epsfxsize=\columnwidth
\centerline{\epsffile{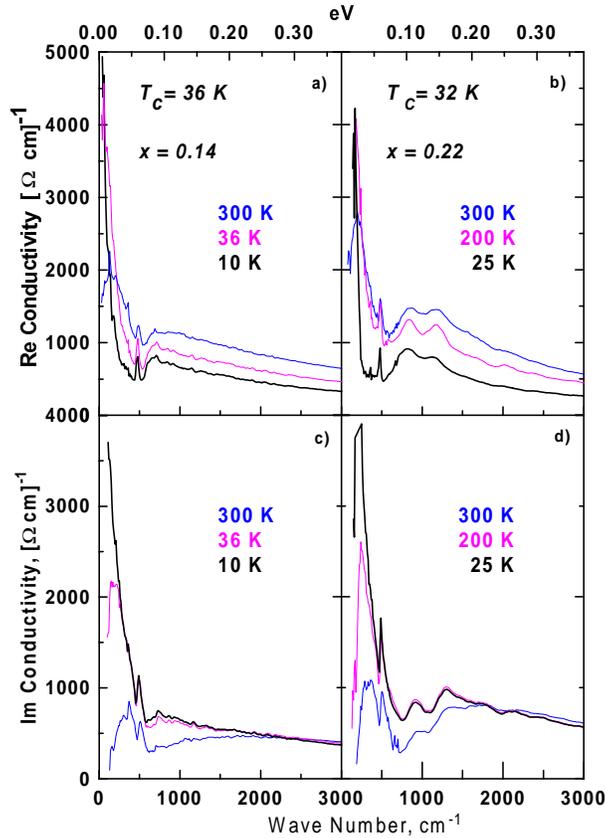}}
\vspace{0.1in}
\caption{The temperature dependence of  the  in-plane  conductivity
of  underdoped La$_{1.86}$Sr$_{0.14}$CuO$_4$ (left  panel) and
overdoped La$_{1.78}$Sr$_{0.22}$CuO$_4$ (right panel) are shown.  The
real part of  conductivity (a,b)  is suppressed below 700 cm$^{-1}$.
The Drude-like peak forms at  room temperature and narrows as T
decreases. The imaginary part of the conductivity
is shown on the lower panel (c,d). Thin line is room
temperature and thick line represents a temperature below
superconducting transition.
}
\label{conductivity}
\end{figure*}

The  frequency dependent scattering rate and the effective mass are 
shown in Fig. 5.  The spectra can conveniently be divided into two 
regions.  In the high frequency region, starting at about  
700~cm$^{-1}$, the scattering rate varies linearly with frequency 
while in the low frequency region there is a clear suppression of  
1/$\tau$($\omega$,T) below this linear trend. We will call this 
frequency region of suppressed scattering the pseudogap state  
region. As the temperature is lowered this suppression becomes 
deeper. The dashed lines in Fig 5 are extrapolations of the high 
frequency linear behaviour to zero frequency. A pseudogap state can 
be defined in terms of this suppression of scattering: the material 
is in the pseudogap state when the scattering rate falls below the 
high frequency straight-line extrapolation. In the low frequency 
limit the scattering rate is proportional to the dc resistivity. Due 
to this, the  1/$\tau$($\omega$,T) suppression should be compared to 
the suppression of $\rho_{DC}$(T) \cite{batlogg} at   temperatures 
below the linear T dependence region. The IR measurement gives us the 
possibility to see both the frequency and the temperature dependence 
of this feature.

The temperature dependence above 700~cm$^{-1}$ is strongly 
influenced by the level of Sr doping. In the underdoped sample 
the high frequency scattering rate is nearly temperature independent. In 
contrast,  the overdoped samples  scattering rate above 700 cm$^{-1}$ 
increases uniformly with temperature.  The curves are seen to be displaced 
parallel to  each other as the temperature is increased. This 
behaviour is also seen in  other overdoped HTSC.\cite{anton_review}

\begin{figure*}%[t]
\leavevmode
\epsfxsize=\columnwidth
\centerline{\epsffile{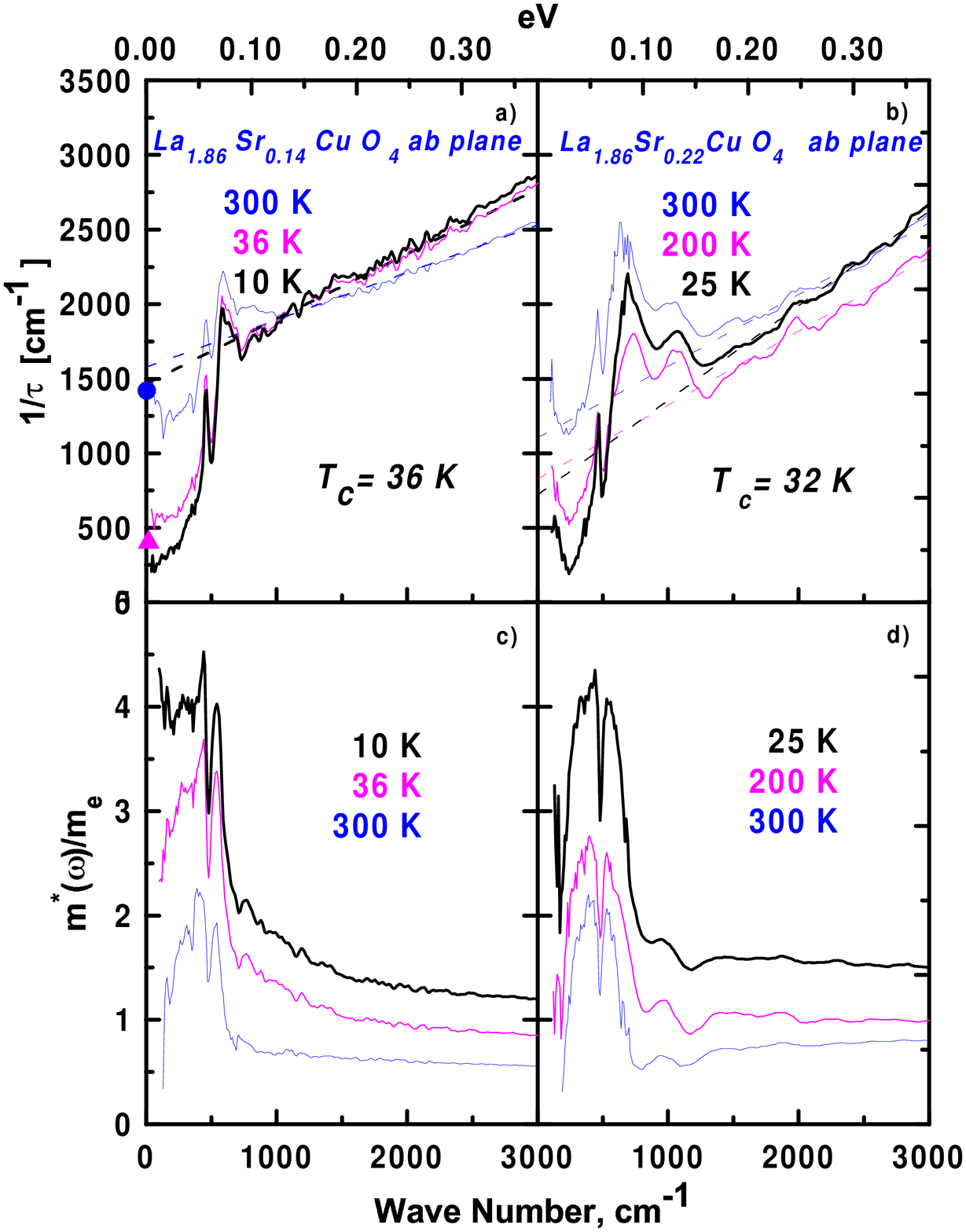}}
\vspace{0.1in}
\caption{Top panel: the frequency dependent scattering rate of
La$_{1.86}$Sr$_{0.14}$CuO$_4$ (a) and La$_{1.78}$Sr$_{0.22}$CuO$_4$
(b) is  calculated using Equation (1). The onset of the suppression
in   a conductivity corresponds to a drastic change in the frequency
dependence of the scattering rate. Above 700~cm$^{-1}$ the scattering
rate is nearly temperature independent and has a linear frequency
dependence in the underdoped sample and temperature dependence in the
overdoped sample. The dashed lines are linear fits to the
scattering rate above 700~cm$^{-1}$.
Below this  frequency the scattering rate varies as $\omega^{1+\delta}$
and shows a strong temperature dependence. Points on $1/\tau$-axis
(a) at $\omega$=0  corresponds to the DC-value calculated from the
resistivity shown in Fig.~1. Bottom panel: The effective mass of
underdoped a) and overdoped b) samples is calculated using Equation
(2). The onset of enhancement of ${m^{\ast}(\omega)\over m_e}$
corresponds to the onset of the suppression of the scattering rate.
The thinnest line corresponds to higher temperature.
}
\label{scattering rate}
\end{figure*}

To emphasize the difference between the underdoped and overdoped regimes 
we fitted the high frequency scattering rate curves with a linear 
function ${1\over\tau(\omega)} = \alpha \omega +\beta$. The 
values of the $\alpha$ and $\beta$ parameters can be found in Table 1.
The zero-frequency intercept, $\beta$, changes only by 7\% in the 
case of underdoped sample but nearly doubles in the overdoped sample
between the superconducting 
transition and 300~K. While $\alpha$ increases with doping, $\beta$ 
is substantially smaller in the overdoped sample.
The behavior of the intercept is consistent with the other HTSC 
materials \cite{anton_review} but not the slope. Puchkov et al. 
found that $\alpha$ decreased with doping in most materials. The  
magnitudes of $\alpha$ and $\beta$ of LSCO seem to be substantially 
lower than those of Y123, Y124, Bi2212 and Tl2202. 
\cite{anton_review}

We found that for La$_{1.78}$Sr$_{0.22}$CuO$_4$ (the overdoped 
sample) $T^{\ast}$ ($\approx$ 300 K) is an order of magnitude higher 
than the superconducting transition temperature $T_c$  (32 K). This  
is significantly different than previous results on overdoped 
cuprates. Theoretical considerations have led to the  suggestion that 
$T^{\ast}$  coincides with the temperature of the superconducting 
transition at the optimal doping level.\cite{lee92,emery&kivelson} 
This seems to apply to Y123 and Y124 where the $T^{\ast}$ and $T_c$ 
curves cross near the optimal doping level, but this is clearly not 
the case here. A detailed examination of the scattering rate curves 
in Fig.~5b suggests that in overdoped sample the suppression of the 
scattering disappears near room temperature implying that $T^{\ast}$ 
for $x=0.22$ is close to 300~K.  This result is consistent with the 
phase diagram based on the data from  transport 
properties.\cite{batlogg} 

In addition to the pseudogap depth and temperature dependence,  
several other features of Figs. 5~a,b should be mentioned. The 
position of the pseudogap remains at 700~cm$^{-1}$ for all 
temperatures. There are also several peaks positioned at 1000 cm$^{-1}$ 
in scattering rate which complicate the analysis, particularly in the 
case of the overdoped samples. These peaks have been attributed to 
polaronic effects.\cite{bipolarons,Thomas92}  

For completeness we plot  the effective 
mass of underdoped sample (Fig.~5c) and the overdoped sample (Fig.~5d). 
As expected,  ${m^{\ast}(\omega)\over m_e}$ rises to a maximum value 
between 2 and 5  in the region of the  
scattering rate suppression. The enhancement of the effective mass in the 
pseudogap state  as well as  the upper limit of 
${m^{\ast}(\omega)\over m_e}$ are similar to what has been 
previously reported for values Y123, Y124 and Bi2212.\cite{anton_review} 

%Table
\vspace{0.1in}
\begin{table}[htpb]
\caption{Linear fit parameters  to the scattering rate of La$_{2-x}$Sr$_{x}$CuO$_4$ }
\begin{tabular}{|c|c|c|c|c|}
Sr content&$T_c$&T&$\alpha$&$\beta$ \\ \hline
&&300~K&0.32&1578\\
x=0.14&36~K &36~K&0.44&1467\\
&&10~K&0.44&1467\\ \hline\
&&300~K&0.39&1260\\
x=0.22& 32~K&200~K&0.5&828\\
&&25~K&0.61&721
\end{tabular}
\end{table}

Before closing we compare our results with data of Gao {\it et 
al.}\cite{Gao} on La$_{2-x}$Sr$_{x}$CuO$_{4+\delta}$ films and 
Quijada {\it et al.}\cite{Quijada} on  oxygen doped 
La$_2$CuO$_{4+\delta}$. Our results in the underdoped case are 
comparable with those of the oxygen doped material, although Quijada 
{\it et al.} did not carry out a frequency dependent scattering rate 
analysis for their underdoped sample.  The film results of Gao {\it 
et al.} are  quite different from our findings.  The films used in that 
study had a strontium level that would 
correspond to optimal doping  in crystals. However, the 
$1/\tau(\omega)$ curves deviate 
markedly from what we observe for  slightly under and overdoped 
samples. The authors performed an extended Drude 
analysis and  found a strongly temperature dependent scattering rate. This is  
in sharp contrast to our results which would suggest a very weak 
temperature dependence. Based on our work, their samples should 
be in the pseudogap state since they have an $x$ value near optimal 
doping. Comparing these results with other systems, in particular 
with Tl2202, two factors suggest the possibility that the films may be 
overdoped. First, their $T_c$ was near 30 K, lower than that 
expected for optimal doping. Secondly, it is known that the oxygen 
level in films can vary substatially and in LSCO oxygen can have a 
major influence on the doping level\cite{zhang}. On the other hand, we 
cannot completely rule out the possibility that all of the crystal 
results are affected by the polishing process, and that the films better
represent  the bulk material. It is clearly important to 
measure films where the oxygen content is controlled by selective 
annealing.

In  conclusion, the optical data in the  far-infrared region, taken 
on two  {\it single-layered} high-$T_c$ superconductors, shows  clear 
evidence of {\it a pseudogap state} in the scattering rate. This 
pseudogap state  extends to  higher temperatures than that observed 
in the multi-layered underdoped cuprates such as YBCO and BSCO. 
Previously, the pseudogap state   feature with $T^{\ast} > T_c$ was 
only observed in the underdoped system. In the case of LSCO this 
feature can be observed in the overdoped regime at a temperature 
substantially above the superconducting transition temperature. This  
suggests that the crossover from the underdoped to    the  overdoped 
regime does not suppress $T^{\ast}$ below $T_c$. The scattering rate  
is similar for both systems in the pseudogap state. At low 
frequencies, $\omega \leq$ 700~cm$ ^{-1}$, the scattering rates are 
{\em temperature} dependent and change with  frequency in a 
{\em non-linear} fashion. Above 700 cm$^{-1}$ this behaviour becomes  
{\em linear}. Within experimental uncertainty the observed high 
frequency scattering rate of the underdoped sample is {\em not 
affected by temperature}.   

% The concluding paragraph should only contain the important points
% not all the points that have been made.

We would like to thank J.D.~Garrett for  help in  aligning the sample 
and also P.C.~Mason, M.~Lumsden and B.D.~Gaulin for  determining  the 
miscut angle of the underdoped LSCO crystal.  We also take this 
opportunity to thank K.C.~Irwin and J.G.~Naeini for the useful 
collaboration. This work was supported by the Natural Sciences and 
Engineering Research  Council of Canada and The Canadian Institute 
for Advanced Research.

% References % \

\end{document}